# Flux Variations of Cosmic Ray Air Showers Detected by LHAASO-KM2A During a Thunderstorm on 10 June 2021*

F. Aharonian[4,5], Q. An(安琪)[6,7], Axikegu(阿西克古)[8], L.X. Bai(白立新)[9], Y.X. Bai(白云翔)[1,3], Y.W. Bao(包逸炜)[10], D. Bastieri[11], X.J. Bi(毕效军)[1,2,3], Y.J. Bi(毕玉江)[1,3], J.T. Cai(蔡金庭)[11], Zhe Cao(曹喆)[6,7], Zhen Cao(曹臻)[1,2,3], J. Chang(常进)[12], J.F. Chang(常劲帆)[1,3,6], E.S. Chen(陈恩生)[1,2,3], Liang Chen(陈良)[1,2,3], Liang Chen(陈亮)[13], Long Chen(陈龙)[8], M.J. Chen(陈明君)[1,3], M.L. Chen(陈玛丽)[1,3,6], S.H. Chen(陈素弘)[1,2,3], S.Z. Chen(陈松战)[1,3], T.L. Chen(陈天禄)[14], X. J. Chen(陈学健)[8†], Y. Chen(陈阳)[10], H.L. Cheng(程皓麟)[2,15,14], N. Cheng(程宁)[1,3], Y.D. Cheng(程耀东)[1,3], S.W. Cui(崔树旺)[16], X.H. Cui(崔晓红)[15], Y.D. Cui(崔昱东)[17], B.Z. Dai(戴本忠)[19], H.L. Dai(代洪亮)[1,3,6], Z.G. Dai(戴子高)[7], Danzengluobu(单增罗布)[14], D. della Volpe[20], K.K. Duan(段凯凯)[12], J.H. Fan(樊军辉)[11], Y.Z. Fan(范一中)[12], Z.X. Fan(范志香)[1,3], J. Fang(方军)[19], K. Fang(方堃)[1,3], C.F. Feng(冯存峰)[21], L. Feng(封莉)[12], S.H. Feng(冯少辉)[1,3], X.T. Feng(丰晓婷)[21], Y.L. Feng(冯有亮)[14], B. Gao(高博)[1,3], C.D. Gao(高川东)[21], L.Q. Gao(高林青)[1,2,3], Q. Gao(高启)[14], W. Gao(高卫)[1,3], W.K. Gao(高伟康)[1,2,3], M.M. Ge(葛茂茂)[19], L.S. Geng(耿利斯)[1,3], G.H. Gong(龚光华)[22], Q.B. Gou(苟全补)[1,3], M.H. Gu(顾旻皓)[1,3,6], F.L. Guo(郭福来)[13], J.G. Guo(郭俊广)[1,2,3], X.L. Guo(郭晓磊)[8], Y.Q. Guo(郭义庆)[1,3], Y.Y. Guo(郭莹莹)[12], Y.A. Han(韩毅昂)[23], H.H. He(何会海)[1,2,3], H.N. He(贺昊宁)[12], S.L. He(何思乐)[11], X.B. He(何新波)[17], Y. He(何钰)[8], M. Heller[20], Y.K. Hor(贺远强)[17], C. Hou(侯超)[1,3], X. Hou(侯贤)[24], H.B. Hu(胡红波)[1,2,3], Q. Hu(胡铨)[7,12], S. Hu(胡森)[9], S.C. Hu(胡世聪)[1,2,3], X.J. Hu(呼晓军)[22], D.H. Huang(黄代绘)[8], W.H. Huang(黄文昊)[21], X.T. Huang(黄性涛)[21], X.Y. Huang(黄晓渊)[12], Y. Huang(黄勇)[1,2,3], Z.C. Huang(黄志成)[8], X.L. Ji(季筱璐)[1,3,6], H.Y. Jia(贾焕玉)[8], K. Jia(贾康)[21], K. Jiang(江琨)[6,7], Z.J. Jiang(姜泽军)[19], M. Jin(金敏)[1,3], M.M. Kang(康明铭)[9], T. Ke(柯通)[1,3], D. Kuleshov[25], B.B. Li(李兵兵)[16], Cheng Li(李澄)[6,7], Cong Li(李骢)[1,3], F. Li(李飞)[1,3,6], H.B. Li(李海波)[1,3], H.C. Li(李会财)[1,3], H.Y. Li(李华阳)[7,12], J. Li(李军)[7,12], Jian Li(李剑)[7], Jie Li(李捷)[1,3,6], K. Li(李凯)[1,3], W.L. Li(李文龙)[21], X.R. Li(李秀荣)[1,3], Xin Li(李昕)[6,7], Xin Li(李新)[8], Y.Z. Li(李一卓)[1,2,3], Zhe Li(李哲)[1,3], Zhuo Li(黎卓)[26], E.W. Liang(梁恩维)[27], Y.F. Liang(梁云峰)[27], S.J. Lin(林苏杰)[17], B. Liu(刘冰)[7], C. Liu(刘成)[1,3], D. Liu(刘栋)[21], H. Liu(刘虎)[8], H.D. Liu(刘海东)[23], J. Liu(刘佳)[1,3], J.L. Liu(刘江来)[28], J.S. Liu(刘佳松)[17], J.Y. Liu(刘金艳)[1,3], M.Y. Liu(刘茂元)[14], R.Y. Liu(柳若愚)[10], S.M. Liu(刘四明)[8], W. Liu(刘伟)[1,3], Y. Liu(刘怡)[11], Y.N. Liu(刘以农)[22], W.J. Long(龙文杰)[8], R. Lu(鲁睿)[19], Q. Luo(罗晴)[16], H.K. Lv(吕洪魁)[1,3], B.Q. Ma(马伯强)[26], L.L. Ma(马玲玲)[1,3], X.H. Ma(马欣华)[1,3], J.R. Mao(毛基荣)[24], A. Masood[8], Z. Min(闵振)[1,3], W. Mitthumsiri[29], Y.C. Nan(南云程)[21], Z.W. Ou(区子维)[17], B.Y. Pang(庞彬宇)[8], P. Pattarakijwanich[29], Z.Y. Pei(裴致远)[11], M.Y. Qi(齐孟尧)[1,3], Y.Q. Qi(祁业情)[16], B.Q. Qiao(乔冰强)[1,3], J.J. Qin(秦家军)[7], D. Ruffolo[29], A. Sáiz[29], C.Y. Shao(邵澄宇)[17], L. Shao(邵琅)[16], O. Shchegolev[25,30], X.D. Sheng(盛祥东)[1,3], J.Y. Shi(石京燕)[1,3], H.C. Song(宋慧超)[26], Yu.V. Stenkin[25,30], V. Stepanov[25], Y. Su(苏扬)[12], Q.N. Sun(孙秦宁)[8], X.N. Sun(孙晓娜)[27], Z.B. Sun(孙志斌)[31], P.H.T. Tam(谭柏轩)[17], Z.B. Tang(唐泽波)[6,7], W.W. Tian(田文武)[2,15], B.D. Wang(王博东)[1,3], C. Wang(王超)[31], H. Wang(王辉)[8], H.G. Wang(王洪光)[11], J.C. Wang(王建成)[24], J.S. Wang(王界双)[28], L.P. Wang(王利苹)[21], L.Y. Wang(王玲玉)[1,3], R. Wang(王冉)[21], R.N. Wang(王润娜)[8], W. Wang(王为)[17], X.G. Wang(王祥高)[27], X.Y. Wang(王祥玉)[10], Y. Wang(王阳)[8], Y.D. Wang(王玉东)[1,3], Y.J. Wang(王岩谨)[1,3], Y.P. Wang(王亚平)[1,2,3], Z.H. Wang(王忠海)[9], Z.X. Wang(王仲翔)[19], Zhen Wang(王振)[28], Zheng Wang(王铮)[1,3,6], D.M. Wei(韦大明)[12], J.J. Wei(魏俊杰)[12], Y.J. Wei(魏永健)[1,2,3], T. Wen(文韬)[19], C.Y. Wu(吴超勇)[1,3ϛ], H.R. Wu(吴含荣)[1,3], S. Wu(武莎)[1,3], X.F. Wu(吴雪峰)[12], Y.S. Wu(吴雨生)[7], S.Q. Xi(席邵强)[1,3], J. Xia(夏捷)[7,12], J.J. Xia(夏君集)[8],

---

*This work is supported in China by National Key R&D program of China under the grants (2018YFA0404201, 2018YFA0404202, 2018YFA0404203, 2018YFA0404204), by NSFC (U2031101, 11475141, 12147208), and in Thailand by RTA6280002 from Thailand Science Research and Innovation.

† X. J. Chen (ChenXJ@my.swjtu.edu.cn)

‡ X. X. Zhou (zhouxx@swjtu.edu.cn)

ϛ C.Y. Wu (wucy@ihep.ac.cn)

G.M. Xiang(项光漫)[2, 13], D.X. Xiao(肖迪泫)[14], G. Xiao(肖刚)[1, 3], G.G. Xin(辛广广)[1, 3], Y.L. Xin(辛玉良)[8], Y. Xing(邢祎)[13], Z. Xiong(熊峥)[1, 2, 3], D.L. Xu(徐东莲)[28], R.X. Xu(徐仁新)[26], L. Xue(薛良)[21], D.H. Yan(闫大海)[24], J.Z. Yan(颜景志)[12], C.W. Yang(杨朝文)[9], F.F. Yang(杨冯帆)[1, 3, 6], H.W. Yang(杨何文)[17], J.Y. Yang(杨佳盈)[17], L.L. Yang(杨莉莉)[17], M.J. Yang(杨明洁)[1, 3], R.Z. Yang(杨睿智)[7], S.B. Yang(杨深邦)[19], Y.H. Yao(姚玉华)[9], Z.G. Yao(姚志国)[1, 3], Y.M. Ye(叶一锰)[22], L.Q. Yin(尹丽巧)[1, 3], N. Yin(尹娜)[21], X.H. You(游晓浩)[1, 3], Z.Y. You(游智勇)[1, 2, 3], Y.H. Yu(于艳红)[7], Q. Yuan(袁强)[12], H. Yue(岳华)[1, 2, 3], H.D. Zeng(曾厚敦)[12], T.X. Zeng(曾婷轩)[1, 3, 6], W. Zeng(曾玮)[19], Z.K. Zeng(曾宗康)[1, 2, 3], M. Zha(查敏)[1, 3], X.X. Zhai(翟徐徐)[1, 3], B.B. Zhang(张彬彬)[10], F. Zhang(张丰)[8], H.M. Zhang(张海明)[10], H.Y. Zhang(张恒英)[1, 3], J.L. Zhang(张建立)[15], L.X. Zhang(张丽霞)[11], Li Zhang(张力)[19], Lu Zhang(张路)[16], P.F. Zhang(张鹏飞)[19], P.P. Zhang(张佩佩)[7, 12], R. Zhang(张瑞)[7, 12], S.B. Zhang(张少博)[2, 15], S.R. Zhang(张少如)[16], S.S. Zhang(张寿山)[1, 3], X. Zhang(张潇)[10], X.P. Zhang(张笑鹏)[1, 3], Y.F. Zhang(张云峰)[8], Y.L. Zhang(张月雷)[1, 3], Yi Zhang(张毅)[1, 12], Yong Zhang(张勇)[1, 3], B. Zhao(赵兵)[8], J. Zhao(赵静)[1, 3], L. Zhao(赵雷)[6, 7], L.Z. Zhao(赵立志)[16], S.P. Zhao(赵世平)[12, 21], F. Zheng(郑福)[31], Y. Zheng(郑应)[8], B. Zhou(周斌)[1, 3], H. Zhou(周浩)[28], J.N. Zhou(周佳能)[13], P. Zhou(周平)[10], R. Zhou(周荣)[9], X.X. Zhou(周勋秀)[8‡], C.G. Zhu(祝成光)[21], F.R. Zhu(祝凤荣)[8], H. Zhu(朱辉)[15], K.J. Zhu(朱科军)[1, 2, 3, 6], and X. Zuo(左雄)[1, 3]

(LHAASO Collaboration)

[1]Key Laboratory of Particle Astrophyics & Experimental Physics Division & Computing Center, Institute of High Energy Physics, Chinese Academy of Sciences, 100049 Beijing, China

[2]University of Chinese Academy of Sciences, 100049 Beijing, China

[3]TIANFU Cosmic Ray Research Center, Chengdu, Sichuan, China

[4]Dublin Institute for Advanced Studies, 31 Fitzwilliam Place, 2 Dublin, Ireland

[5]Max-Planck-Institut for Nuclear Physics, P.O. Box 103980, 69029 Heidelberg, Germany

[6]State Key Laboratory of Particle Detection and Electronics, China

[7]University of Science and Technology of China, 230026 Hefei, Anhui, China

[8]School of Physical Science and Technology & School of Information Science and Technology, Southwest Jiaotong University, 610031 Chengdu, Sichuan, China

[9]College of Physics, Sichuan University, 610065 Chengdu, Sichuan, China

[10]School of Astronomy and Space Science, Nanjing University, 210023 Nanjing, Jiangsu, China

[11]Center for Astrophysics, Guangzhou University, 510006 Guangzhou, Guangdong, China

[12]Key Laboratory of Dark Matter and Space Astronomy & Key Laboratory of Radio Astronomy, Purple Mountain Observatory, Chinese Academy of Sciences, 210023 Nanjing, Jiangsu, China

[13]Key Laboratory for Research in Galaxies and Cosmology, Shanghai Astronomical Observatory, Chinese Academy of Sciences, 200030 Shanghai, China

[14]Key Laboratory of Cosmic Rays (Tibet University), Ministry of Education, 850000 Lhasa, Tibet, China

[15]National Astronomical Observatories, Chinese Academy of Sciences, 100101 Beijing, China

[16]Hebei Normal University, 050024 Shijiazhuang, Hebei, China

[17]School of Physics and Astronomy (Zhuhai) & School of Physics (Guangzhou) & Sino-French Institute of Nuclear Engineering and Technology (Zhuhai), Sun Yat-sen University, 519000 Zhuhai & 510275 Guangzhou, Guangdong, China

[18]Dipartimento di Fisica dell'Università di Napoli "Federico II", Complesso Universitario di Monte Sant'Angelo, via Cinthia, 80126 Napoli, Italy.

[19]School of Physics and Astronomy, Yunnan University, 650091 Kunming, Yunnan, China

[20]Département de Physique Nucléaire et Corpusculaire, Faculté de Sciences, Université de Genève, 24 Quai Ernest Ansermet, 1211 Geneva, Switzerland

[21]Institute of Frontier and Interdisciplinary Science, Shandong University, 266237 Qingdao, Shandong, China

[22]Department of Engineering Physics, Tsinghua University, 100084 Beijing, China

[23]School of Physics and Microelectronics, Zhengzhou University, 450001 Zhengzhou, Henan, China

[24]Yunnan Observatories, Chinese Academy of Sciences, 650216 Kunming, Yunnan, China

[25]Institute for Nuclear Research of Russian Academy of Sciences, 117312 Moscow, Russia

[26]School of Physics, Peking University, 100871 Beijing, China

[27]School of Physical Science and Technology, Guangxi University, 530004 Nanning, Guangxi, China

[28]Tsung-Dao Lee Institute & School of Physics and Astronomy, Shanghai Jiao Tong University, 200240 Shanghai, China

[29]Department of Physics, Faculty of Science, Mahidol University, 10400 Bangkok, Thailand

[30]Moscow Institute of Physics and Technology, 141700 Moscow, Russia

[31]National Space Science Center, Chinese Academy of Sciences, 100190 Beijing, China

## Abstract

The Large High Altitude Air Shower Observatory (LHAASO) has three sub-arrays, KM2A, WCDA and WFCTA. The flux variations of cosmic ray air showers were studied by analyzing the KM2A data during the thunderstorm on 10 June 2021. The number of shower events that meet the trigger conditions increases significantly in atmospheric electric fields, with maximum fractional increase of 20%. The variations of trigger rates (increases or decreases) are found to be strongly dependent on the primary zenith angle. The flux of secondary particles increases significantly, following a similar trend with that of the shower events. To better understand the observed behavior, Monte Carlo simulations are performed with CORSIKA and G4KM2A (a code based on GEANT4). We find that the experimental data (in saturated negative fields) are in good agreement with simulations, assuming the presence of a uniform electric field of -700 V/cm with a thickness of 1500 m in the atmosphere above the observation level. Due to the acceleration/deceleration by the atmospheric electric field, the number of secondary particles with energy above the detector threshold is modified, resulting in the changes in shower detection rate.

## 1. Introduction

The correlation between the cosmic ray variations and thunderstorm electric fields has been a hot topic in the interdisciplinary science of cosmic ray physics and atmospheric physics. During thunderstorms, the intensity of atmospheric electric fields (AEFs) could be up to 2000 V/cm [1-3], and the polarity can change dramatically [4]. By acceleration/deceleration of strong electric fields, the secondary charged particles in extensive air showers (EAS) could be significantly affected.

In 1924, the concept of "runaway electrons" was first suggested by Wilson [5], and then developed by Gurevich et al. in 1992 [6]. A secondary electron (with tiny mass) in an EAS can be accelerated by the strong AEF in thunderclouds, and gain energy exceeding that lost in ionization and bremsstrahlung. The acceleration of secondary electrons there can be runaway, which may ionize more air molecules, generating even more electrons, which are further accelerated by the field, resulting in an avalanche. This process is now commonly called the relativistic runaway electron avalanche (RREA) [7]. Building on the RREA theory, Dwyer et al. [8–11] presented the relativistic feedback mechanism, including X-ray feedback (important for stronger AEF) and positron feedback (dominating at lower field strengths).

Over the years, several high-altitude experiments, i.e., the Carpet air shower array [12], EAS-TOP [13], ASEC [14, 15], Tibet ASγ [16], ARGO-YBJ [17], SEVAN at Lomnický Štít [18, 19], a network of thermal neutron detectors [20], and detectors on Mount Norikura [21, 22] and Mount Fuji [23], have reported the flux variations of cosmic rays associated with thunderstorm episodes. The RREA mechanism could be the source of the strong thunderstorm ground enhancements (TGEs), where the particle flux measured at the ground level exceeds the background values by several times. Several satellite-based experiments, such as CGRO [24], AGILE [25] and Fermi-GBM [26], have observed thousands of terrestrial gamma-ray flashes (TGFs), sub-millisecond gamma-ray emissions originated from bremsstrahlung by runaway electrons. The RREA process is thought to be responsible for TGFs.

According to the proposed theory [8, 27], the AEF strength threshold ($E_{th}$) required to generate the RREA process is found to be a function of the altitude. At the altitude of the LHAASO site, $E_{th}$≈1660 V/cm. Since the field strength required to trigger the RREA process is very large, it should only occur rarely. Some ground-based experiments observed smaller TGEs with the flux increases less than 10% [15, 18, 28]. Several teams reported the decreases of secondary particle intensity during thunderstorms [12, 17, 29]. These phenomena cannot be reasonably explained by the RREA process. By performing Monte Carlo simulations, it was found [30] that the flux of the secondary positions and electrons decreases in a certain range of positive fields (a positive AEF is defined as the direction pointing towards the ground) and increases in negative fields. This phenomenon was attributed to the asymmetry in number and energy of electrons and positrons in EAS. This hypothesis was supported by the experimental results of ARGO-YBJ in scaler mode and simulated development of electrons and positrons in EAS [17].

In addition, there have been some reports on the energy and lateral distribution of secondary particles during thunderstorms. The effect of the electric fields on the energy of secondary particles was analyzed [31-33] and it was found that the energy spectrum is softened in the presence of the field. The variation of the lateral density of secondary positrons and electrons was simulated, and it was shown that the lateral distribution becomes wider [34].

During a thunderstorm, the intensity and the polarity of AEF may vary dramatically, mostly caused by the discharge of lightning. Lightning is a common geophysical phenomenon, characterized as a very long electrical spark and usually divided into two categories, cloud-to-ground (CG) lightning and intra-cloud (IC) lightning [35, 36].

The correlation between lightning and the cosmic ray variations has also been studied. Numerous TGF-related lightning events have been observed by satellites and it is widely believed that IC lightning triggers the TGF events [37-39]. However, according to the ground-based experiments, terminations of TGEs by lightning have been observed [40, 41].

Even though many insights about the correlation between cosmic rays and thunderstorms are obtained thanks to the works by many authors, because of the complexity of thunderstorms, some questions still remain unresolved. For example, does the IC lightning trigger or terminate the enhancement of cosmic rays? Up to now, there is no complete image of flux variations of cosmic rays from different incident directions during thunderstorms. These issues are still far from being understood completely, and more observations and simulations are required to shed light on them.

LHAASO is suitable to study the correlation between thunderstorms and the variations of cosmic rays, thanks to its large active area and its location at high altitude with frequent thunderstorms. In this work, the variations measured by KM2A during the thunderstorm on 10 June 2021 are studied in detail. Comparing the data to the Monte Carlo simulations, a simple model of the AEF was obtained.

## 2. The LHAASO-KM2A Detector

LHAASO, a new generation hybrid extensive air shower (EAS) array, has been constructed at Haizi Mountain (4410 m a. s. l., in Daocheng, Sichuan province, China), aiming for the studies of cosmic ray physics and γ-ray astronomy [42, 43]. It consists of three types of detector arrays: a 1.3 $km^2$ array (KM2A), a 78000 $m^2$ water cherenkov detector array (WCDA), and a wide field-of-view air cherenkov telescope array (WFCTA) with 18 telecopes. As the largest component of LHAASO, KM2A contains 5216 Electromagnetic particle Detectors (EDs) and 1188 Muon Detectors (MDs). The ED array covers an area of 1.3 km$^2$ in a triangular grid. It is divided into two parts: a central part with 4911 EDs (15 m spacing) in a circular area with radius of 575 m, and a guard ring with 305 EDs (30 m spacing) surrounding the central area with outer radius of 635 m. The EDs mainly detect the electromagnetic particles in the shower, which are used to reconstruct the shower parameters (i.e., core position, arrival direction and primary energy). The 1188 MDs are deployed in a triangular grid with spacing of 30 m in the central part of the array. Enclosed within a cylindrical concrete tank, the whole MD detector is covered by a soil layer of 2.5 m thickness to absorb the secondary positrons, electrons and γ-rays in showers, but not the muons. More details about the designs of ED and MD can be found elsewhere [43].

The KM2A detectors were constructed and merged into the data acquisition system (DAQ) in stages. The first 33 EDs started operating in February 2018. The half of the KM2A array, including 2365 EDs and 578 MDs, started running from 27 December 2019. Through the analysis of the 136 live days of data detected by the 1/2 KM2A array, the first observation of the Crab Nebula and the detector performance were presented [44]. The 3/4 KM2A array (including 3978 EDs and 917 MDs) began data taking on 1 December 2020. At last, the whole array began stably running from 20 July 2021. By analyzing the partly completed KM2A data, some important results, such as PeVatrons from Galactic sources, were obtained [45].

For some ground-based experiments, there are two independent data acquisition systems, corresponding to the shower and scaler operation modes [17, 46]. In KM2A, the scaler mode [47] is under study. In this work, data of the shower mode were analyzed. In the shower mode, the trigger logic requires at least 20 EDs fired within a time window of 400 ns. For each shower event, the DAQ records 10 μs of data from all EDs and MDs that have signals over the thresholds. The ED signals are used for the shower reconstruction, while the MD signals are used to select γ-ray induced showers [44].

In order to study the cosmic ray variations during thunderstorms, a ground-based electric field mill (Boltek EFM-100) was installed at LHAASO observatory on 17 September 2019. The EFM-100 is designed to measure the AEF and provide a real-time electric field versus time graph. With the help of the time rate of change of the step electric field change, the lightning distance is estimated. The EFM-100 can record lightning out to about 48 km, and the location accuracy was ~1.6 km [48]. To prevent the rotor from being damaged, the mill was located on the roof of the WCDA-2 building. Considering the AEF enhancement due to the location above the ground level, the reference field mill was mounted flush with the surface of the ground to calibrate. After the correction factor was applied, the dynamic range of the AEF measurement is from -270 V/cm to 270 V/cm.

## 3. Data Selection

Thunderstorms are common weather phenomena at high altitudes. From October 2019 to March 2022, there were more than 200 thunderstorm events detected by the field mill at LHAASO observatory. Most of them were recorded from April to September. A thunderstorm that occurred on 10 June 2021 (here we call it Thunderstorm 20210610) is most notable. The specific information is presented in Fig.1. Thunderstorm 20210610 is a complex episode, which is characterized by frequent changes in electric field polarity, a long time of field strength in saturation, and the presence of 11 nearby (<1.6 km) lightning strikes. In this work, the cosmic-ray variations detected by KM2A during Thunderstorm 20210610 were studied in detail.

From Fig.1, it can be seen that the AEF disturbance lasted about two and a half hours, from 10:15:36 to 12:42:48 (UT) on 10 June 2021. Much of the time, the field mill was saturated. During this interval, the AEF polarity changed as many as 46 times. There were 1712 lightning strikes estimated by EFM-100. Most of them were distant lightning strikes with a mean distance of about 27 km. For ground-based experiments, if the thundercloud is far from the detector, the cosmic ray variations are small due to the heavy atmospheric attenuation [49]. Here, we only consider the effects of near-earth AEF and nearby lightning strikes on the cosmic rays measured by KM2A. The 24 lightning strikes with distances less than 10 km are also shown in Fig. 1.

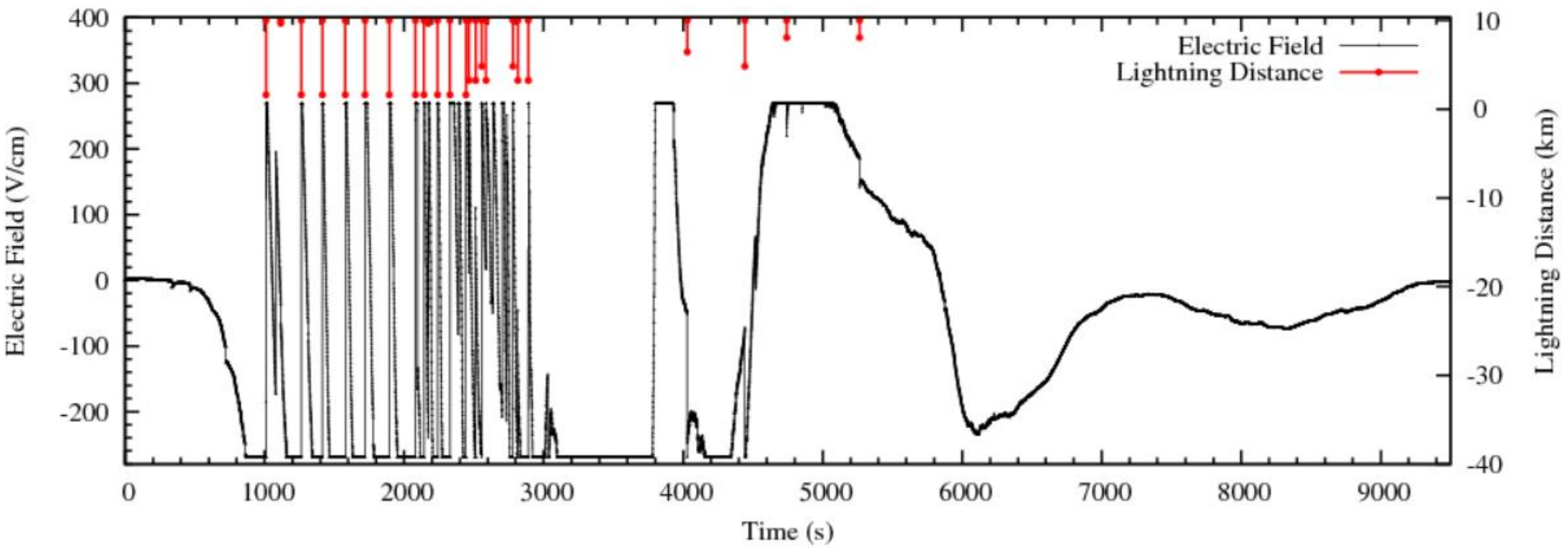

Fig. 1.Variation of the near-earth AEF and the distance to lightning strike (<10 km) recorded by Boltek EFM-100. Time zero of the *x*-axis is 10:10:00 UT (the same

To ensure the validity of our results, the operational status of detectors was carefully checked from different angles (such as the distributions of anode/dynode charge of particles for each detector, the distributions of hit number for each detector as a function of time, etc). Some detectors were powered off due to strong lightning. We found that about 8% of the detectors were not working properly after 10:57:06. Only data from detectors that were operating normally before 10:57:06 are used in this work. They are shown in Fig. 2.

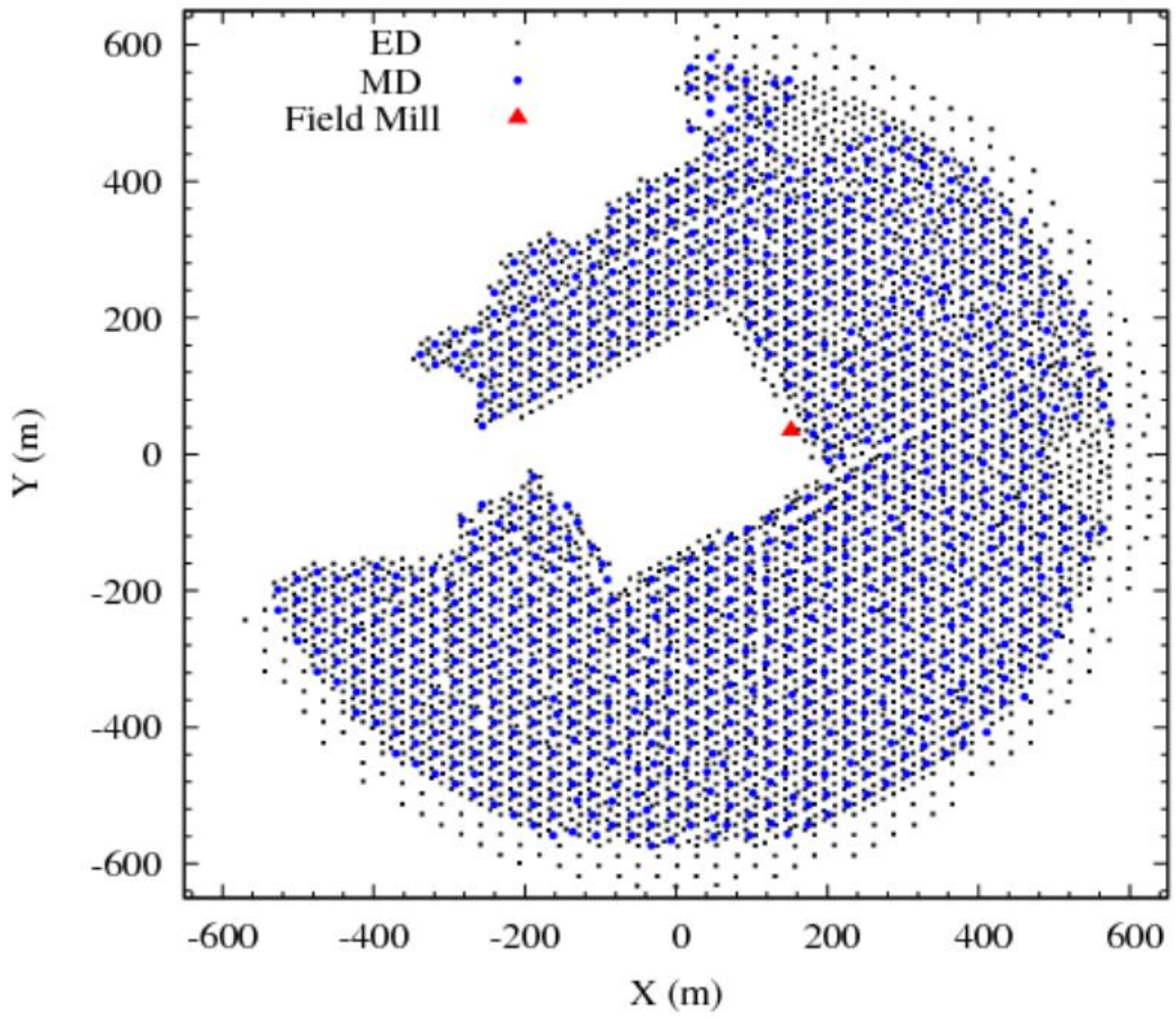

Fig. 2. The layout of the 3/4 KM2A array with detectors that were operating normally during Thunderstorm 20210610. The black squares and blue dots indicate the EDs and MDs, respectively. The red solid triangle indicates the position of Boltek EFM-100.

## 4. Observational Results

Analyzing the data detected by KM2A, the flux variations of cosmic ray shower events and ground secondary particles during Thunderstorm 20210610 are as follows.

### 4.1 The flux variations of shower events in KM2A

During thunderstorms, the secondary particles in EAS are strongly affected by the AEFs. As a result, the number of shower events that meet the KM2A trigger conditions (at least 20 fired EDs within 400 ns) will also change. During Thunderstorm 20210610, the distributions of AEF, nearby lightning distance and shower rate detected by KM2A are shown in Fig. 3.

From Fig. 3 (a), it can be seen that the thunderstorm event is very complex. From 720 s to 2826 s (after 10:10:00 UT), the polarity and strength of the fields changed sharply. During the thunderstorm, the absolute values of AEF intensity exceeded the measuring range of the field mill several times. The total time in saturated status is 165 s in positive AEF and 918 s in negative AEF. There were 11 nearby lightning strikes (with distances less than 1.6 km) measured by the Boltek EFM-100 during these 24 minutes. Based on the variations of AEFs, these lightning strikes are most likely negative CG flashes. The thunderclouds can be simply assumed to be dipolar [50]. The main negative charge region is distributed at the lower dipole and the positive one is located at the top of thundercloud. If the negatively charged area is close to the ground, a large amount of positive charge will be induced on the ground. When the induced charge is sufficient, strong negative CG flashes will be initiated. As a result, the lightning will destroy the lower dipole (negative charge), and positive charges at the top of the thundercloud are exposed, resulting in the AEF value jumping from negative saturation to positive saturation [51]. Within 10-50 s, the lower negative charge can recover.

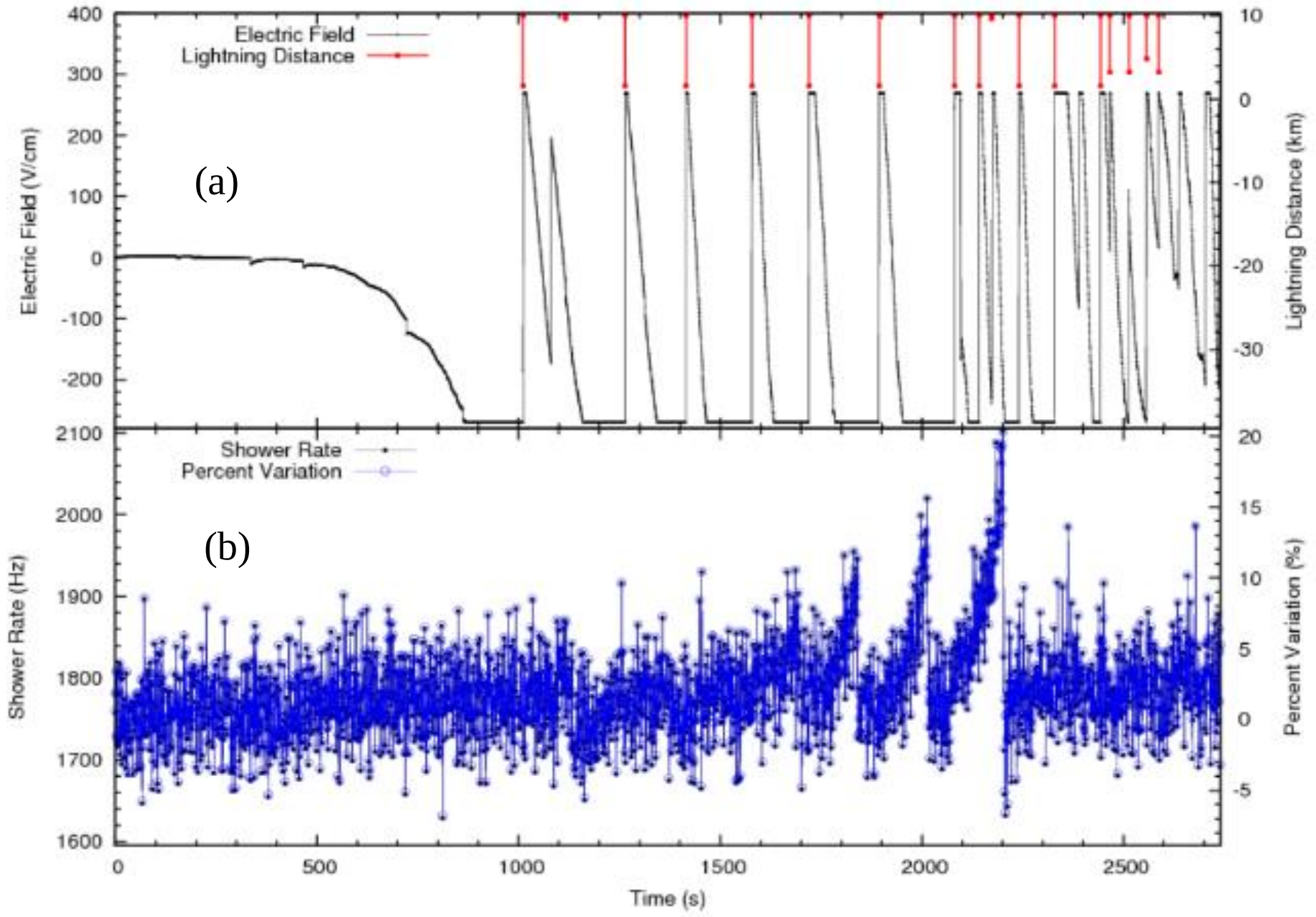

Fig. 3. Variations of AEF, the distance to lightning strike (a) and shower rate (b) per second during Thunderstorm 20210610.

With respect to the shower rate measured in a period of 2000 s before the thunderstorm (defined as fair weather), the percent variations are calculated. From Fig. 3 (b), the shower rate significantly increases in thunderstorm fields, with maximum increase exceeding 20%. Due to the acceleration by AEFs, the secondary particles with energy exceeding the detector threshold increases, and then more shower events satisfy the trigger conditions, resulting in the increase in the shower rate.

From the studies in references [30, 52], the AEFs have different effects on cosmic rays with different zenith angles ($\theta$). By analyzing the reconstructed events in KM2A (the details about event reconstruction can be found elsewhere [44]), the variations of shower rate in different zenith angle ranges are shown in Fig. 4. In this work, the events with reconstructed zenith angles less than 60$^{o}$ are analyzed. From 820 s to 2202 s, the trigger rate present structural increases in lower zenith angle ranges ($0^{o}<\theta\leq30^{o}$), with the maximum exceeding 29% at 2201 s. Whereas for higher zenith angle ranges ($30^{o}<\theta\leq60^{o}$), it decreases by up to 18%. From Fig. 4, we can clearly see the opposite variation structures. As a result, the total variations of shower rates are reduced (see in Fig. 3 (b)).

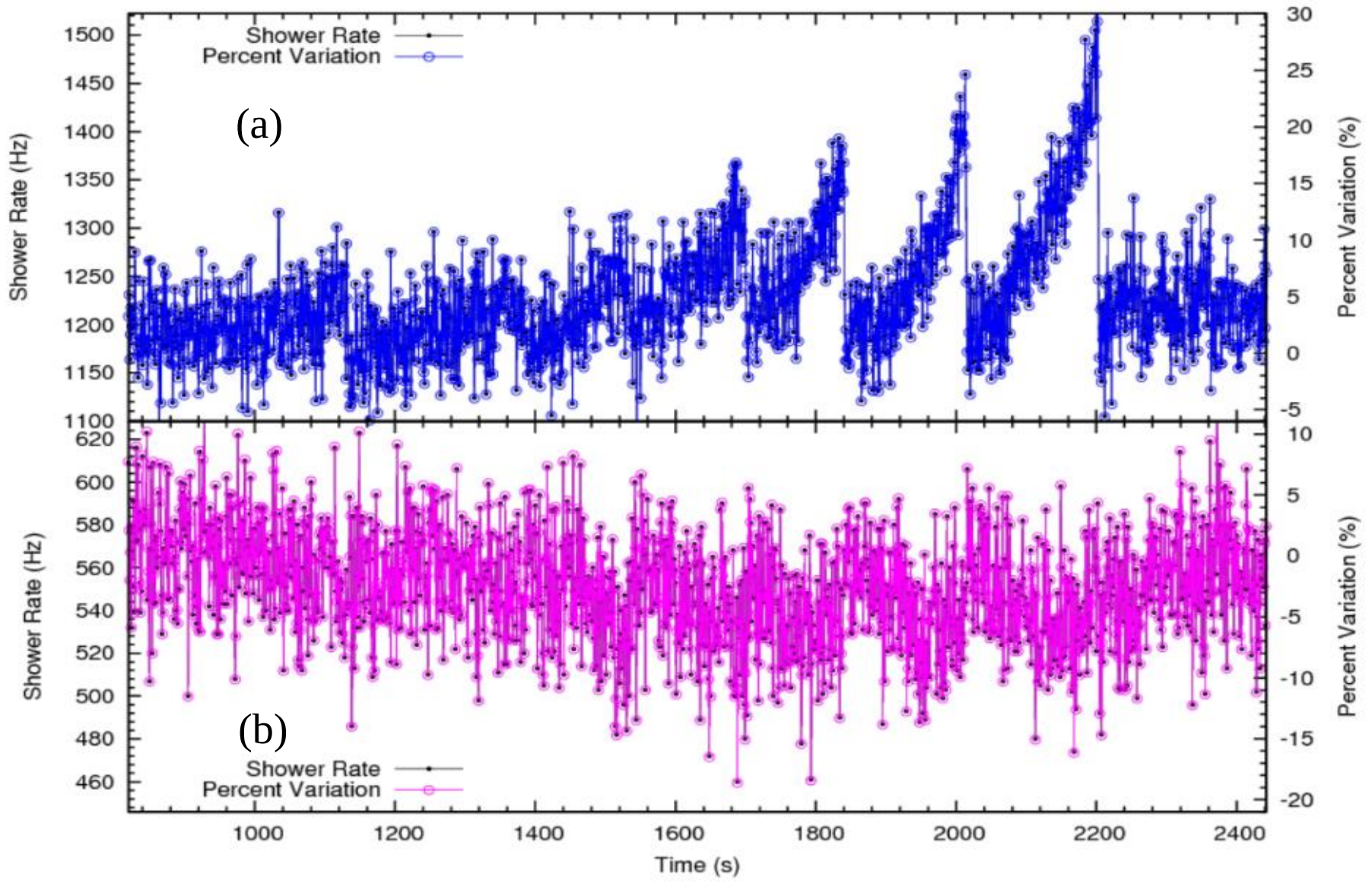

Fig. 4. The shower rate variations per second in zenith angle ranges of 0-30° (a) and 30-60° (b), respectively.

Analyzing the data in saturated negative fields, the average value of shower rate as a function of zenith angle (4°/bin) is shown in Fig. 5. The result in fair weather is also shown for comparison. In saturated negative fields, the rate increases in smaller zenith angle ranges, and decreases in larger zenith angle ranges.

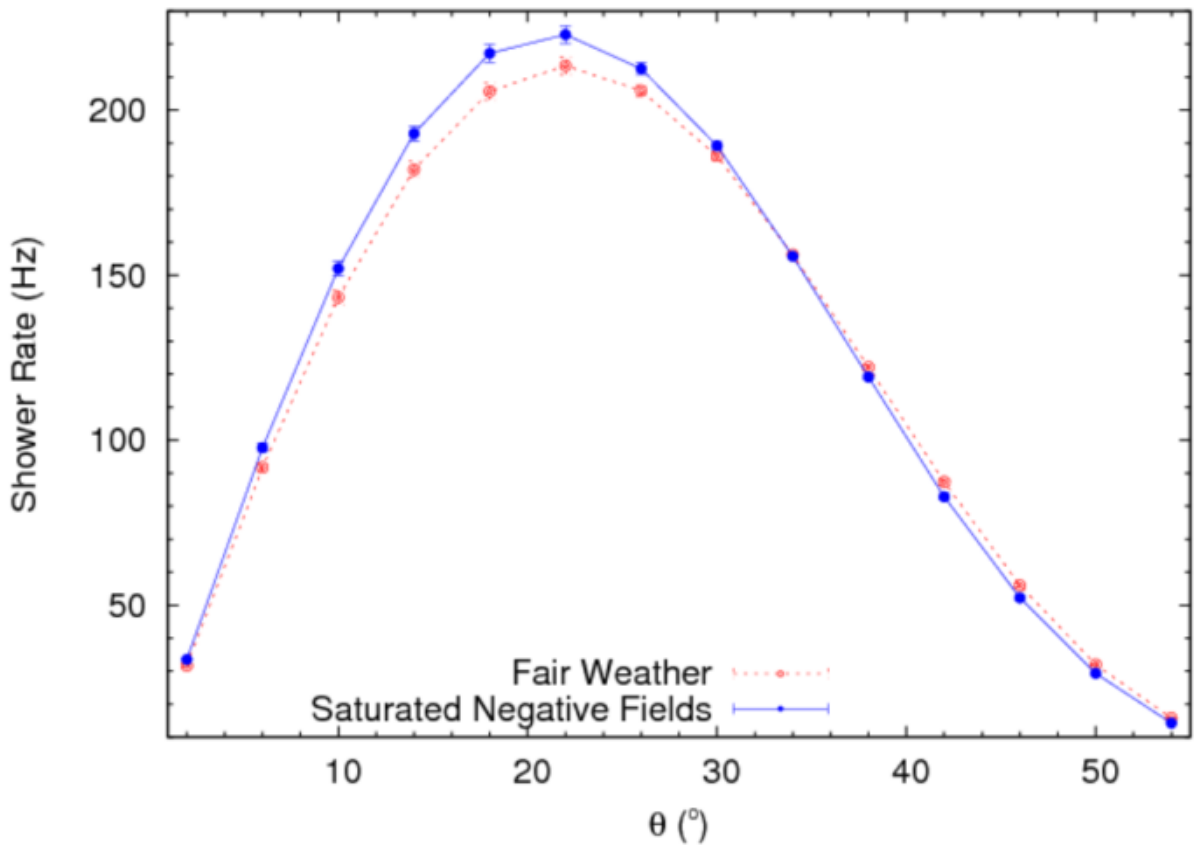

Fig. 5. The shower rates of the zenith angle distributions in saturated negative fields.
The results in fair weather are plotted for comparison.

With respect to the average value measured in fair weather, the percent variation of shower rate in saturated negative fields is shown in Fig. 6. The rate increases if the zenith angle is small, with maximum amplitude 6.5%. However, with the increase of the zenith angle, the increase becomes smaller and smaller. It starts declining at zenith angle larger than 33°, to the maximum change of about -10%.

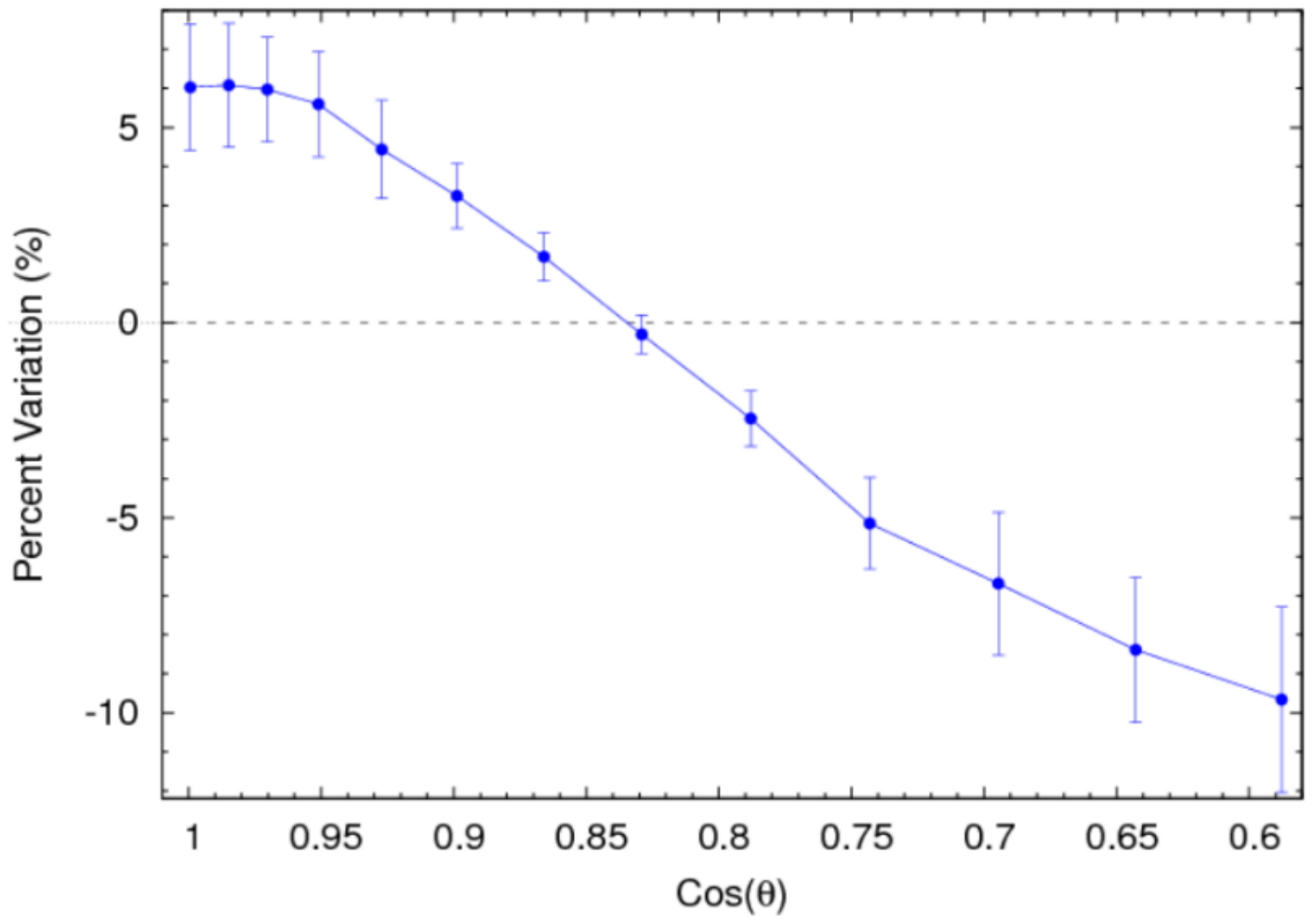

Fig. 6. The average variation of the zenith angle distribution in saturated negative fields.

## 4.2 The flux variations of ground-level secondary particles in KM2A

The KM2A array uses trigger conditions based on the particles recorded by the ED array [44]. To understand the changes in shower rate in AEFs, detailed studies on the flux variations of ground-level secondary particles are necessary.

For each triggered event, the DAQ records the data from all EDs and MDs that have signals over the thresholds. Variation of the average number of particles per shower recorded by EDs ($N_e$) and the average number of muons per shower recorded by MDs ($N_\mu$) are shown in Fig. 7. The variation in AEF is also shown in Fig. 7.

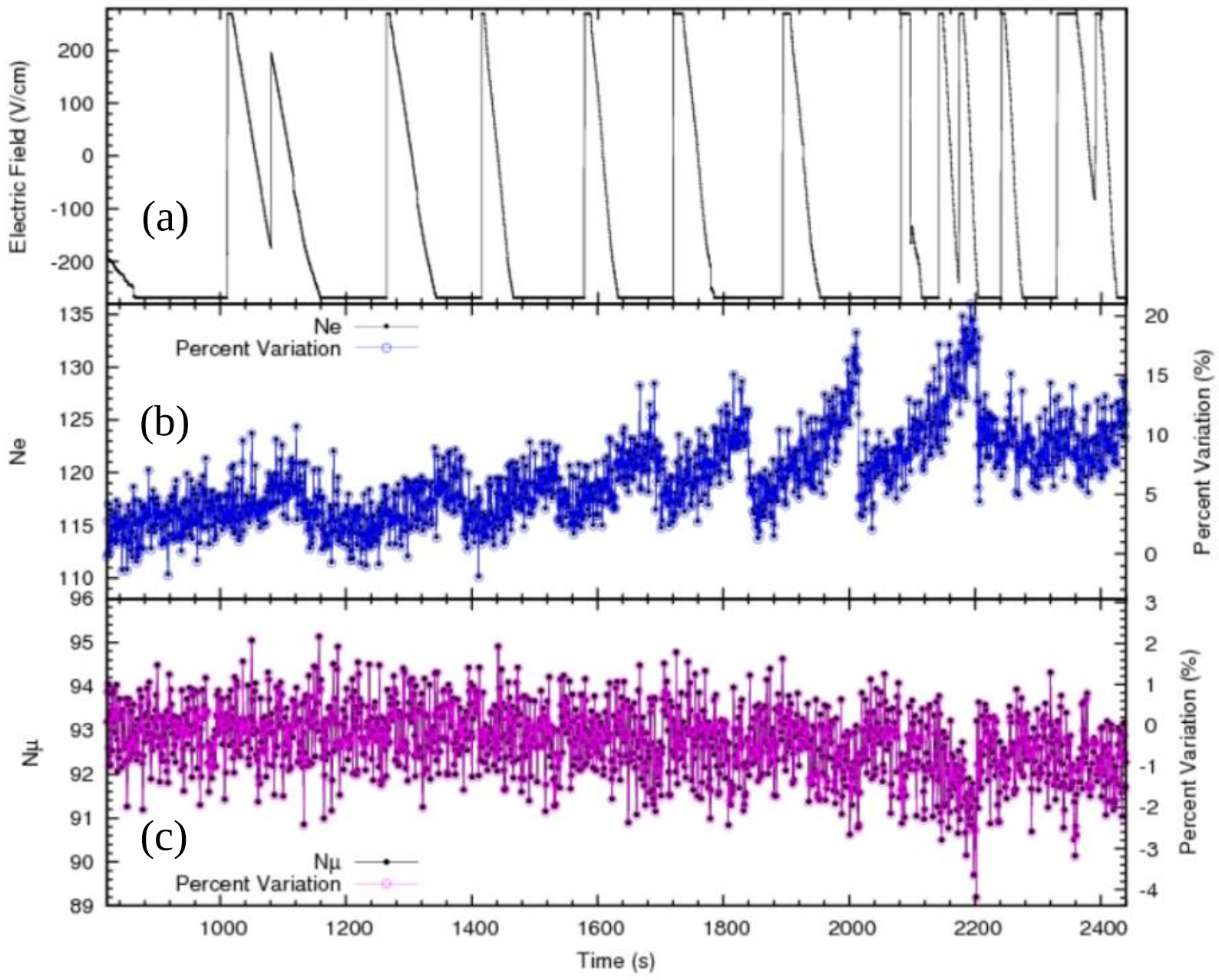

Fig. 7. Variations of AEFs (a), $N_e$ (b) and $N_\mu$ (c) per second during Thunderstorm 20210610.

From Fig. 7 (b), we can see a significant increase in the number of particles detected by the ED array in thunderstorm fields. As shown in Fig. 7 (c), the MD array shows decreases in count rate. Due to the acceleration of the secondary charged particles when they cross the layers of AEF, the number of ground-level secondary particles with energy above the detector threshold will increase [17, 30]. At the same time, the mean energies of secondary positrons/electrons are much lower than that of muons [33]. Most muons detected by MDs have energies about 1 GeV [43]. According to the Bethe's theory, if the energy is greater than ~1.4 MeV, the drag force increases with the particle energy [53]. This means that the AEF has more effects on particles with smaller energies, i.e., on positrons and electrons, but has small effects on muons with larger energies [30]. As a result, a clear increase in particle count rate per shower is observed in the ED array. From Fig. 7 (c), it can be seen that the muon rate per shower shows no increases, but rather decreases in the MD array. This is most likely due to the enhanced rate of shower events (with lower primary energy and fewer secondary muons) when there are strong fields (see Fig. 3). As a result, the average muon number per shower shows a decrease.

During a thunderstorm, the noise trigger recorded by the detector will increase. To study the field effects on the flux variations of ground-level secondary particles detected by ED, the noise during thunderstorm needs to be considered. For each shower event, the trigger time is set at 0, and the DAQ records all hits within 5000 ns before or after the trigger time [44]. According to the trigger logic, the data were divided into 2 parts. The hits ($N_{off}$) between -5000 ns and -500 ns are mostly noise, and the hits ($N_{on}$) from -500 ns to 5000 ns are mostly signal. From $N_{on}$ and $N_{off}$, the secondary particles ($N_s$) from EAS can be calculated by the formula:

$$N_s = N_{on} - 1.22 \cdot N_{off}\,, \tag{1}$$

where 1.22=5500/4500 is the ratio between the widths of time windows.

The distributions of AEFs, $N_{on}$, $N_{off}$ and $N_s$ per second in the ED array are shown in Fig. 8. With respect to the rate measured in fair weather, the calculated percent variations are also shown in Fig. 8. It can be seen that the particle numbers in both time windows are clearly increasing. From Fig. 8 (c), it can be seen that the electronic noise increases and the increased number of $N_{off}$ is less than 7. From Fig.8 (d), after removing the electronic noise, $N_s$ still shows a significant increase in strong negative fields, with a maximum value up to 20%. In positive fields, the variation of $N_s$ is not significant.

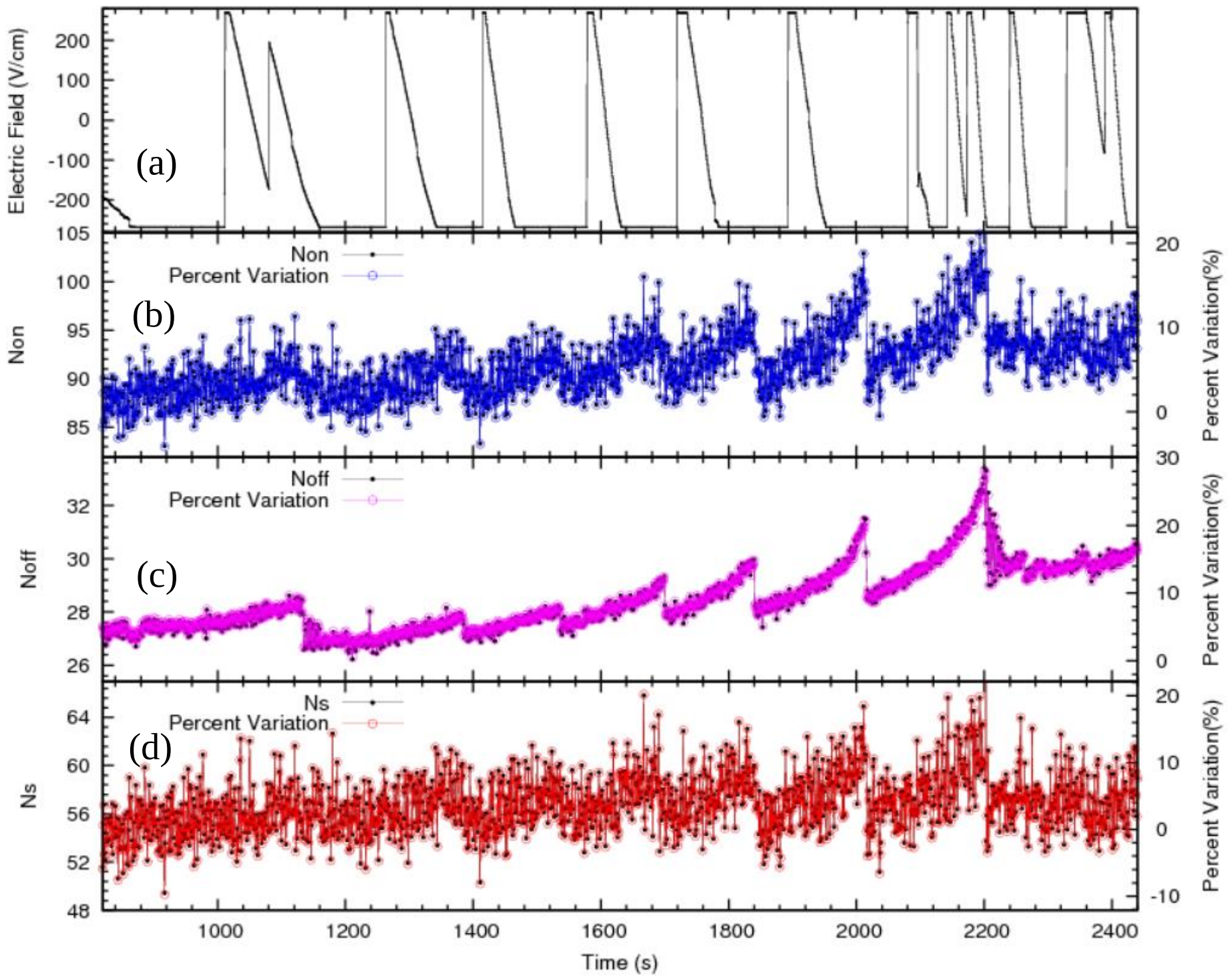

Fig. 8. Variations of AEFs (a), $N_{\mathrm{on}}$ (b), $N_{\mathrm{off}}$ (c) and $N_{\mathrm{s}}$ (d) per second during Thunderstorm 20210610.

From 820 s to 2202 s (see Fig. 8), the ED count rate presents structural changes, and 7 peaks appear at times of negative AEF. After reaching the maximum value, it drops abruptly (mostly within 4-12 s), and then starts to increase again. Why did the particle flux decrease rapidly? According to some previous observations [40, 41], lightning may terminate the flux increase. In references [17, 30], the variation of cosmic ray fluxes is related to the intensity and thickness of AEF during thunderstorms. Unfortunately, the Boltek EFM-100 was in a saturation state during the sharp drops, so the information of AEF changes and lightning strikes was not recorded. That means it is impossible to analyze their effects on the particles directly in this work. It should be noted that another electric field mill with a wider dynamic range of ±1000 V/cm was installed at LHAASO observatory on 27 Oct. 2021. In the future work, more thunderstorm events should be analyzed with detailed information on electric fields and lightning strikes.

## 5. Simulation results and discussion

To understand the data recorded by KM2A, the EAS development and the detector response is simulated.

The CORSIKA code (version 7.7410) [54] is used to simulate air showers in the atmosphere. QGSJET (Quark Gluon String model with JETs) and GHEISHA (Gamma Hadron Electron Interaction SHower code) are programs developed to describe hadronic interactions. In this work, the selected hadronic interaction models are QGSJETII-04 in high energy range (> 80 GeV) and GHEISHA in low energy range (< 80 GeV). In view of the acceleration by the field, the energy threshold of the secondary particles (positrons, electrons and photons) has been set to the lowest possible value, 50 keV. This means that secondary particles below this energy are discarded before they may be accelerated by the AEF to higher energies. As a result, the AEF effects may be underestimated. The horizontal and vertical strength of the geomagnetic field components used in simulations are $B_X = 34.8\ \mu$T and $B_Z = 36.2\ \mu$T, calculated by the IGRF model of Earth's magnetic field [55]. We assume proton primaries with arrival direction evenly distributed in the sky, with a zenith angle less than 60°. According to the energy threshold of the KM2A [43, 44], the simulated primary particles are selected as protons with energy ranging from 1 to $10^5$ TeV following a power-law function with a spectral index of −2.7.

To simulate the KM2A detector response, a specific software, G4KM2A [56], was developed in the framework of the GEANT4 package [57]. In these simulations, at least 20 fired EDs within 400 ns are required for a shower trigger, as in the observations.

To study the AEF effects, a simple model is used, with a vertical and uniform AEF in a layer of atmosphere. Here, the presence of a negative AEF is assumed. Fig. 9 shows the variations of the triggered shower events measured by 3/4 KM2A array for different AEF intensities, with the layer thickness of 1000 m (extending from the detector altitude of 4410 m up to 5410 m). It can be seen that the flux of shower events triggered by KM2A increases with the AEF strength. In a field of -1200 V/cm, the rate of shower events is enhanced by up to 20%.

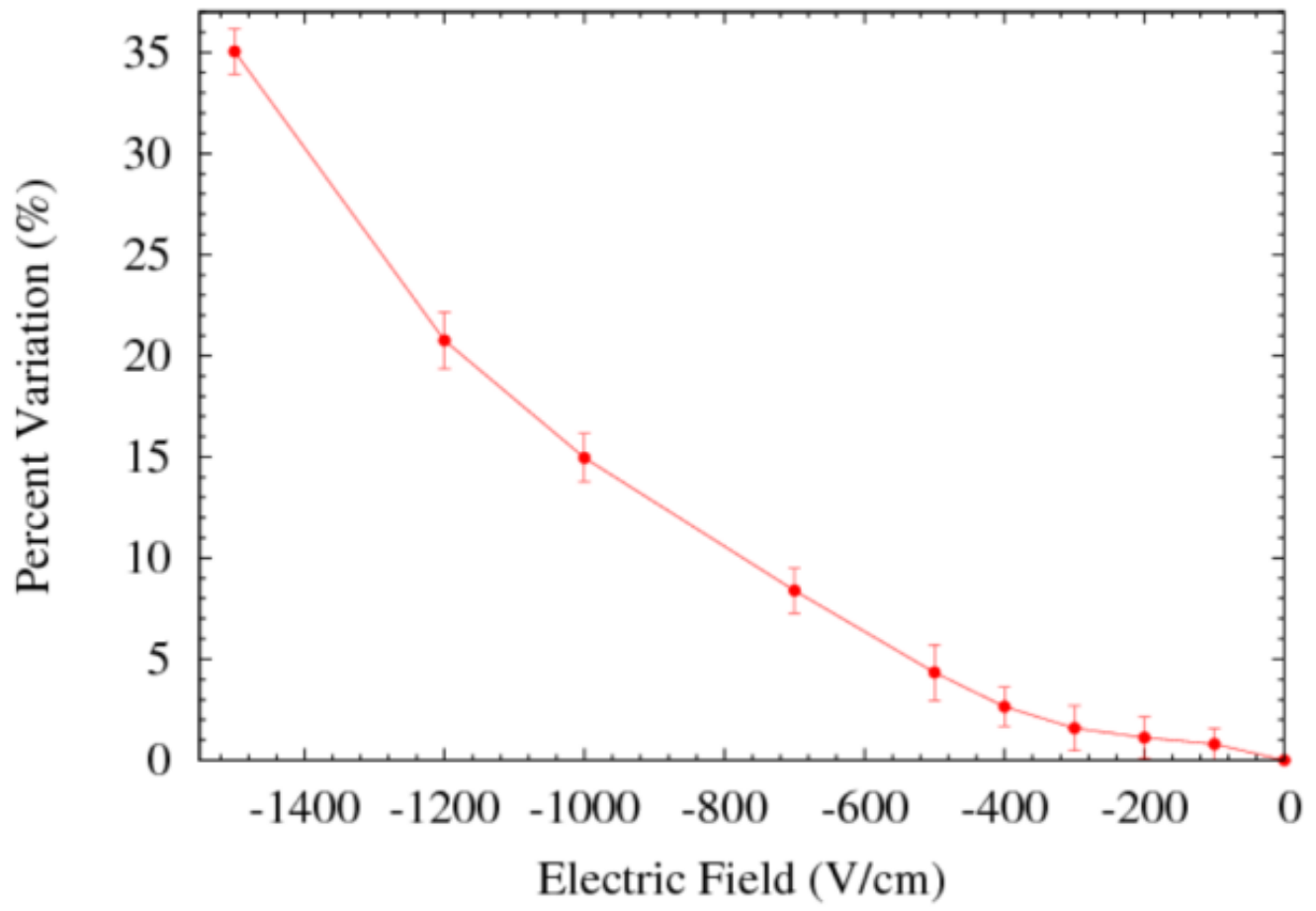

Fig. 9. Simulations: Percent variations of the shower event rate as a function of the electric field intensity with AEF layer thickness of 1000 m.

Fig. 10 shows the flux variations as a function of the thickness of the field in the atmosphere above the detector, assuming an AEF intensity of -1000 V/cm. We can see that the trigger event rate dramatically increases at small thickness, and then the curve flattens out when the thickness is higher than 1000 m. That means the AEF at higher altitudes has a small influence on the development of EAS.

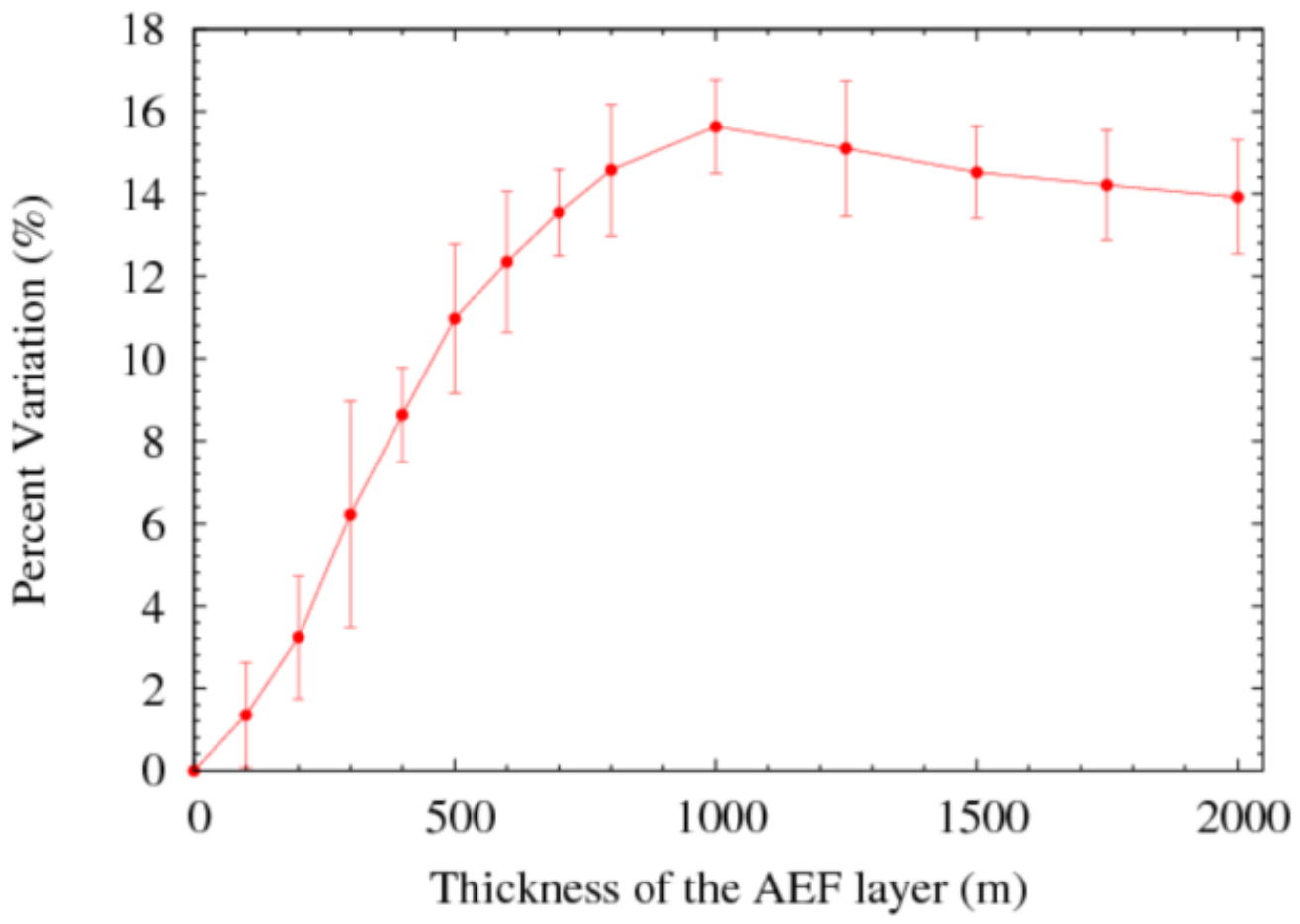

Fig. 10. Simulations: Percent variations of the shower event rate as a function of the thickness of a -1000 V/cm AEF layer.

From Fig. 9 and Fig. 10, it can be seen that the variations of shower rate are related to the strength and thickness of the AEF layer. Comparing to the KM2A data, the range of saturated electric field is mostly from -1200 to -400 V/cm, and its spatial extent above the observation level is greater than 200 m.

To compare the average effects in saturated electric fields during Thunderstorm 20210610, we tried several values of AEF strengths and thicknesses. The results show that in a field of -700 V/cm, with a layer thickness of 1500 m, the simulations agree well with data. The comparison of simulations with experimental data is shown in Fig.11.The number of shower events increases for smaller zenith angles, but decreases for higher zenith angles. We can see that the simulated trends are consistent with the observed data.

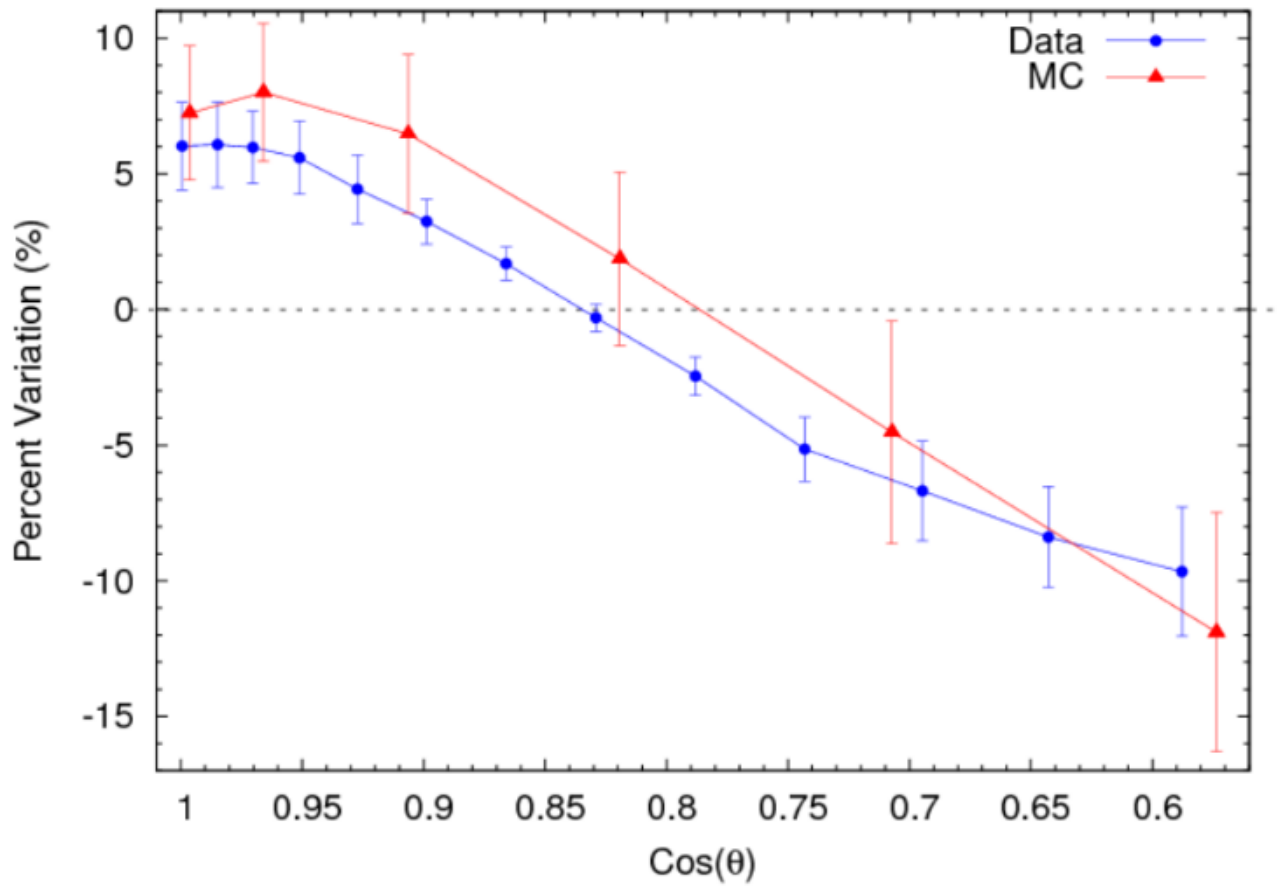

Fig. 11. The shower event rate variations as a function of zenith angle in a field of -700 V/cm with a layer thickness of 1500 m, compared to experimental data.

## 6. Summary

In this work, the cosmic ray variations measured by KM2A during Thunderstorm 20210610 are studied in detail. Significant changes of the shower events and ground-level secondary particles are found and explained.

During the thunderstorm, the shower rate detected by KM2A shows a clear increase, with percentage variation exceeding 20%. The variations of shower rate are related to the zenith angle. For smaller zenith angle ranges, the average shower rate increases by up to 6.5% in saturated negative fields. For larger zenith angles, the increase becomes smaller and smaller. At a certain value of zenith angle, the shower rate begins to decrease down to about -10%.

The flux increases of ground-level secondary particles have been observed during times of strong AEF, with the maximum enhancements of 20% (after considering the effects of electronic noise). During the thunderstorm, the secondary particles increase with a special pattern, alternating between gradual increases and rapid drops.

The Monte Carlo simulations have been performed by using CORSIKA and G4KM2A. The flux of trigger events in the 3/4 KM2A array increases with the strength and thickness of AEFs. Assuming a uniform AEF of -700 V/cm with a thickness of 1500 m, our simulations are consistent with the experimental data in saturated negative fields.

Our data can be understood in terms of acceleration/deceleration effects of the AEF on charged particles in the air showers. During thunderstorm, the number of secondary particles hitting the detector is change, leading to the shower rate change.

### Acknowledgments

The authors would like to thank all staff members who work at the LHAASO site at 4400 meters above sea level year round to maintain the detector and keep the electrical power supply and other components of the experiment operating smoothly. We are grateful to Chengdu Management Committee of Tianfu New Area for the constant financial support for research with LHAASO data.